\documentclass{elsart}

\usepackage{amssymb,amsmath}

\newcommand\rf[1]{(\ref{eq:#1})}
\newcommand\lab[1]{\label{eq:#1}}
\newcommand\nonu{\nonumber}
\newcommand\br{\begin{eqnarray}}
\newcommand\er{\end{eqnarray}}
\newcommand\be{\begin{equation}}
\newcommand\ee{\end{equation}}

\newcommand\lb{\lbrack}
\newcommand\rb{\rbrack}

\newcommand\llb{\left\lbrack}
\newcommand\rrb{\right\rbrack}

\renewcommand\({\left(}
\renewcommand\){\right)}
\newcommand\bv{\bigm\vert}               
\newcommand\bgv{\bigg\vert}              

\newcommand\bc{\begin{center}}
\newcommand\ec{\end{center}}

\newcommand\partder[2]{\frac{{\partial {#1}}}{{\partial {#2}}}}

\newcommand\funcder[2]{{{\delta {#1}}\over{\delta {#2}}}}

\renewcommand\b{\beta}

\renewcommand\d{\delta}

\newcommand\eps{\epsilon}
\newcommand\vareps{\varepsilon}
\newcommand\g{\gamma}
\newcommand\G{\Gamma}

\newcommand\h{\frac{1}{2}}
\renewcommand\k{\kappa}
\renewcommand\l{\lambda}
\renewcommand\L{\Lambda}
\newcommand\m{\mu}
\newcommand\n{\nu}

\newcommand\vp{\varphi}
\renewcommand\P{\Phi}
\newcommand\pa{\partial}

\newcommand\pr{\prime}

\newcommand\s{\sigma}

\renewcommand\t{\tau}
\renewcommand\th{\theta}

\newcommand\wti{\widetilde}

\newcommand\cA{{\mathcal A}}

\newcommand\cF{{\mathcal F}}

\newcommand\cP{{\mathcal P}}

\newcommand{\ct}[1]{\cite{#1}}
\newcommand{\bib}[1]{\bibitem{#1}}

\newcommand\PRL[3]{\textsl{Phys. Rev. Lett.} \textbf{#1} (#2) #3}

\newcommand\PRD[3]{\textsl{Phys. Rev.} \textbf{D#1} (#2) #3}

\newcommand\PLB[3]{\textsl{Phys. Lett.} \textbf{#1B} (#2) #3}
\newcommand\CQG[3]{\textsl{Class. Quantum Grav.} \textbf{#1} (#2) #3}

\newcommand\AoP[3]{\textsl{Ann. of Phys.} \textbf{#1} (#2) #3}

\newcommand\IJMPA[3]{\textsl{Int. J. Mod. Phys.} \textbf{A#1} (#2) #3}

\newcommand\Xdot{\stackrel{.}{X}}

\newcommand\rdot{\stackrel{.}{r}}

\begin{document}

\begin{frontmatter}



\title{Einstein-Rosen ``Bridge'' Needs Lightlike Brane Source}


\author[BGU]{Eduardo Guendelman\corauthref{cor1}},
\ead{guendel@bgu.ac.il}
\corauth[cor1]{Corresponding author -- tel. +972-8-647-2508, fax +972-8-647-2904.}
\author[BGU]{Alexander Kaganovich\corauthref{cor1}}
\ead{alexk@bgu.ac.il}
\address[BGU]{Department of Physics, Ben-Gurion University of the Negev,
P.O.Box 653, IL-84105 ~Beer-Sheva, Israel}

\author[INRNE]{Emil Nissimov},
\ead{nissimov@inrne.bas.bg}
\author[INRNE]{Svetlana Pacheva}
\ead{svetlana@inrne.bas.bg}
\address[INRNE]{Institute for Nuclear Research and Nuclear Energy, Bulgarian Academy
of Sciences, Boul. Tsarigradsko Chausee 72, BG-1784 ~Sofia, Bulgaria}

\begin{abstract}
The Einstein-Rosen ``bridge'' wormhole solution proposed in the classic
paper \ct{einstein-rosen} {\em does not} satisfy the vacuum Einstein equations
at the wormhole throat. We show that the fully consistent formulation of the original
Einstein-Rosen ``bridge'' requires solving Einstein equations of bulk $D=4$
gravity coupled to a lightlike brane with a well-defined world-volume action.
The non-vanishing contribution of Einstein-Rosen ``bridge'' solution to the
right-hand side of Einstein equations at the throat matches precisely the surface
stress-energy tensor of the lightlike brane which automatically occupies the
throat (``horizon straddling'') -- a feature triggered by the world-volume
lightlike brane dynamics.
\end{abstract}

\end{frontmatter}

\section{Introduction}

The Einstein-Rosen ``bridge'' space-time proposed in the classical paper
\ct{einstein-rosen} is historically one of the first examples of what
later became known as wormhole space-time manifolds (for a review, see
\ct{visser-book,WH-rev} and references therein).

In a series of recent papers \ct{our-WH,rot-WH} we have explored the novel
possibility of employing {\em lightlike branes} (\textsl{LL-branes} for short)
as natural self-consistent gravitational sources for traversable
wormhole space-times, in other words, generating wormhole solutions in
self-consistent bulk gravity-matter systems coupled to
\textsl{LL-branes} through dynamically derived world-volume \textsl{LL-brane}
stress-energy tensors. Namely, we have provided in \ct{our-WH,rot-WH}
a systematic general scheme to
construct self-consistent spherically symmetric or rotating cylindrical wormhole
solutions via \textsl{LL-branes}, such that the latter occupy the wormhole throats
and match together two copies of exterior regions of spherically symmetric or
rotating cylindrical black holes (the regions beyond the outer horizons).
These wormhole solutions combine the features of the original Einstein-Rosen
``bridge'' manifold \ct{einstein-rosen} (wormhole throat located at horizon)
with the feature ``charge without charge'' of Misner-Wheeler
wormholes \ct{misner-wheeler}. They have been also shown to be {\em traversable}
w.r.t. the {\em proper time} of travelling observers \ct{rot-WH}.

There exist several other types of physically interesting wormhole solutions
in the literature generated by different types of matter and without horizons.
For a recent discussion, see ref.\ct{lemos-bronnikov} and references therein.

As a particular case of our construction in \ct{our-WH,rot-WH}, the matching of
two exterior regions of Schwarzschild space-time at the horizon surface $r=2m$
through an \textsl{LL-brane} turns out to be the self-consistent realization of the
original Einstein-Rosen ``bridge''. Namely, the Einstein-Rosen ``bridge''
metric in its original form from \ct{einstein-rosen} is {\em not} a solution
of the vacuum Einstein equations but rather it requires the presence of an
\textsl{LL-brane} source at $r=2m$ -- a feature not recognized in the original
Einstein-Rosen work \ct{einstein-rosen}. It is the main purpose of the
present note to explain the latter in more detail.

Let us particularly emphasize that here and in what follows we consider the
Einstein-Rosen ``bridge'' in  its original formulation in ref.\ct{einstein-rosen}
as a four-dimensional space-time manifold consisting of two copies of the
exterior Schwarz\-schild space-time region matched along the horizon.
On the other hand, the nomenclature of ``Einstein-Rosen bridge'' in several standard
textbooks (\textsl{e.g.} \ct{MTW}) uses the Kruskal-Szekeres manifold and it
is {\em not equivalent} to the original construction in \ct{einstein-rosen}. Namely,
the two regions in Kruskal-Szekeres space-time corresponding to the outer
Schwarzschild space-time region ($r>2m$) and labeled $(I)$ and $(III)$ in \ct{MTW}
are generally
{\em disconnected} and share only a two-sphere (the angular part) as a common border
($U=0, V=0$ in Kruskal-Szekeres coordinates), whereas in the original Einstein-Rosen
``bridge'' construction the boundary between the two identical copies of the
outer Schwarz\-schild space-time region ($r>2m$) is a three-dimensional hypersurface
\mbox{($r=2m)$}.

In what follows we will make an essential use of the explicit world-volume
Lagrangian formalism for \textsl{LL-branes} proposed earlier in
refs.\ct{LL-brane-main}.
There are several characteristic features of \textsl{LL-branes} which drastically
distinguish them from ordinary Nambu-Goto branes:

(i) They describe intrinsically lightlike modes, whereas Nambu-Goto branes describe
massive ones.

(ii) The tension of the \textsl{LL-brane} arises as an {\em additional
dynamical degree of freedom}, whereas Nambu-Goto brane tension is a given
{\em ad hoc} constant. 
This is an important feature significantly distinguishing our \textsl{LL-brane}
models from the previously proposed {\em tensionless} $p$-branes (for a review,
see ref.\ct{lindstroem-etal}) which rather resemble a $p$-dimensional continuous
distribution of massless point-particles. 

(iii) Consistency of \textsl{LL-brane} dynamics in a spherically or axially
symmetric gravitational background of codimension one requires the presence
of an event horizon which is automatically occupied by the \textsl{LL-brane}
(``horizon straddling'' according to the terminology of ref.\ct{Barrabes-Israel}).

(iv) When the \textsl{LL-brane} moves as a {\em test} brane in spherically or
axially symmetric gravitational backgrounds its dynamical tension exhibits
exponential ``inflation/deflation'' time behaviour \ct{inflation-all}
-- an effect similar to the ``mass inflation'' effect around black hole horizons
\ct{israel-poisson}.

Let us also note that \textsl{LL-branes} by themselves play an important role in
the description of various physically important phenomena in general relativity,
such as impulsive lightlike signals arising in cataclysmic astrophysical events
\ct{barrabes-hogan}, the ``membrane paradigm'' \ct{membrane-paradigm} of black hole
physics and the thin-wall approach to domain walls coupled to
gravity \ct{Israel-66,Barrabes-Israel,Dray-Hooft} (see also \ct{BGG}).

In Section 2 below we show that the original Einstein-Rosen ``bridge'' metric
fails to satisfy the vacuum Einstein equations due to the appearance on the
r.h.s. of a non-vanishing ill-defined (as distribution) $\d$-function singularity
at the throat. This indicates the presence of some kind of matter source
concentrated on the throat -- a three-dimensional lightlike hypersurface
connecting the two ``universes'' carrying the geometry of the exterior
Schwarzschild space-time region.

In Section 3 we briefly review the main properties of the \textsl{LL-brane}
world-volume Lagrangian dynamics which are of utmost importance for our main
construction in the following Section 4.
We propose there a different metric describing the
Einstein-Rosen ``bridge'' manifold, which satisfy the Einstein equations with
a well-defined $\d$-function contribution on the r.h.s. identified as the
stress-energy tensor of an \textsl{LL-brane} coupled to bulk gravity. As
mentioned above, the latter construction is a particular case of the general
construction of spherically and rotating cylindrical wormholes via
\textsl{LL-branes} proposed in \ct{our-WH,rot-WH}.

In Section 5 we provide an alternative construction of Einstein-Rosen ``bridge''
space-time as a {\em lightlike} limit of spherically symmetric wormhole with a
{\em timelike} ``thin shell'' at the throat. In particular, we explain the
reason why we are not encountering any divergencies in the lightlike limit
when joining two exterior Schwarzschild regions along timelike ``thin shell''
with our choice of coordinates unlike the case with the standard treatment
using Gaussian normal coordinates (see \textsl{e.g.} \ct{visser-book};
the latter formalism was specifically designed for timelike ``thin shells'',
without having in mind a lightlike limit).
For an alternative approach appropriate also for
lightlike ``thin shells'', see refs.\ct{Barrabes-Israel,poisson-kit}. The
principal difference w.r.t. the present formalism (refs.\ct{LL-brane-main}
and Section 3 below) is that in our case the \textsl{LL-brane} dynamics is
systematically derived from world-volume action principle.

\section{Einstein-Rosen ``Bridge'' Fails To Satisfy Vacuum Einstein Equations}

Let us start with the coordinate system proposed in \ct{einstein-rosen},
which is obtained from the original Schwarzschild coordinates by defining
$u^2=r-2m$, so that the Schwarzschild metric becomes:
\be
ds^2 = - \frac{u^2}{u^2 + 2m} (dt)^2 + 4 (u^2 + 2m)(du)^2 +
(u^2 + 2m)^2 \( (d\th)^2 + \sin^2 \th \,(d\vp)^2\) \; .
\lab{E-R-metric}
\ee
Then Einstein and Rosen take two identical copies of the exterior Schwarzschild
space-time region ($r>2m$) by letting the new coordinate $u$ to vary between
$-\infty$ and $+\infty$
(\textsl{i.e.}, we have the same $r\geq 2m$ for $\pm u$). The two
Schwarzschild exterior space-time regions must be matched at the horizon $u=0$.

Let us examine whether the original Einstein-Rosen solution satisfy the
vacuum Einstein equations everywhere. To this end let us consider the Levi-Civita
identity (see \textsl{e.g.} \ct{frankel}):
\be
R^0_0 = - \frac{1}{\sqrt{-g_{00}}} \nabla^2 \(\sqrt{-g_{00}}\)
\lab{levi-civita-id}
\ee
valid for any metric of the form
$ds^2 = g_{00} (r) (dt)^2 + h_{ij}(r,\th,\vp) dx^i dx^j$ and where $\nabla^2$
is the three-dimensional Laplace-Beltrami operator
$\nabla^2 =\frac{1}{\sqrt{h}}\partder{}{x^i}\(\sqrt{h}\,h^{ij}\partder{}{x^j}\)$.
The Einstein-Rosen metric \rf{E-R-metric} solves $R^0_0 = 0$ for all $u\neq 0$.
However, since $\sqrt{-g_{00}} \sim |u|$ as $u \to 0$ and since
$\frac{\pa^2}{{\pa u}^2} |u| = 2 \d (u)$, Eq.\rf{levi-civita-id} tells us that:
\be
R^0_0 \sim \frac{1}{|u|} \d (u) \sim \d (u^2) \; ,
\lab{ricci-delta}
\ee
and similarly for the scalar curvature $R \sim \frac{1}{|u|} \d (u) \sim \d (u^2)$.
From \rf{ricci-delta} we conclude that:

(i) The non-vanishing r.h.s. of \rf{ricci-delta} exhibits the explicit presence of
some lightlike matter source on the throat -- an observation which is missing in
the original formulation \ct{einstein-rosen} of the Einstein-Rosen ``bridge''.
In fact, the problem with the metric \rf{E-R-metric} satisfying the vacuum Einstein
equations at $u=0$ has been noticed in ref.\ct{einstein-rosen}, where in Eq.(3a)
the authors multiply Ricci tensor by an appropriate power of the determinant $g$
of the metric \rf{E-R-metric} vanishing at $u=0$ so as to enforce
fulfillment of the vacuum Einstein equations everywhere, including at $u=0$.

(ii) The coordinate $u$ in \rf{E-R-metric} is {\em inadequate} 
for description of the original Einstein-Rosen ``bridge'' at the throat due to
the {\em ill-definiteness} as distribution of the r.h.s. in \rf{ricci-delta}.

We will now describe an alternative construction of the Einstein-Rosen ``bridge''
wormhole as a spherically symmetric wormhole with Schwarzschild geometry
produced via \textsl{LL-brane} sitting at its throat in a self-consistent formulation,
namely, solving Einstein equations with a surface stress-energy tensor
of the lightlike brane derived from a well-defined world-volume brane action.
Moreover, we will show that the mass parameter $m$ of the Einstein-Rosen ``bridge''
is not a free parameter but rather is a function of the {\em dynamical}
\textsl{LL-brane} tension.

To this end we will employ the Finkelstein-Eddington coordinates for the
Schwarzschild metric \ct{EFM} (see also \ct{MTW}):
\be
ds^2 = - A(r) (dv)^2 + 2 dv\,dr + r^2 \llb (d\th)^2 + \sin^2\th (d\vp)^2\rrb
\quad ;\quad A(r) = 1 - \frac{2m}{r} \; .
\lab{EF-metric}
\ee
The advantage of the metric \rf{EF-metric} over the metric in standard
Schwarzschild coordinates is that both \rf{EF-metric} as well as the
corresponding Christoffel coefficients {\em do not} exhibit coordinate
singularities on the horizon $(r=2m)$.

Let us introduce the following modification of \rf{EF-metric}:
\be
ds^2 = - {\wti A} (\eta) (dv)^2 + 2 dv\,d\eta +
{\wti r}^2(\eta) \llb (d\th)^2 + \sin^2\th (d\vp)^2\rrb \; ,
\lab{our-EF-metric}
\ee
where we substituted $r$ with a new coordinate $\eta$ via $r=2m +|\eta|$,
\textsl{i.e.}:
\be
{\wti A} (\eta) = A (2m + |\eta|) = \frac{|\eta|}{|\eta| + 2m} \quad ,\quad
{\wti r}(\eta) = 2m + |\eta| \; .
\lab{our-EF-metric-coeff}
\ee
The metric describes two identical copies of Schwarzschild {\em exterior}
space-time region ($r > 2m$), 
which correspond to
$\eta >0$ and $\eta <0$, respectively, and which are ``glued'' together at the
horizon $\eta = 0$ (\textsl{i.e.}, $r=2m$), where the latter will serve as a throat
of the overall wormhole solution. This is precisely the space-time manifold
of the original Einstein-Rosen ``bridge'' construction (cf. Eq.\rf{E-R-metric})
in terms of the Eddington-Finkelstein coordinate system.

As we will see in what follows, the appropriate coordinate to describe the full
Einstein-Rosen ``bridge'' manifold (including at the wormhole throat) is precisely
$\eta$ rather than the original Einstein-Rosen's coordinate $u$. Obviously, the metric
\rf{our-EF-metric}--\rf{our-EF-metric-coeff} is smooth everywhere except at the
horizon $\eta = 0$ where it is only continuous but not differentiable.
Therefore, it is clear that the pertinent Ricci tensor and the scalar curvature
will exhibit well-defined distributional contributions $\sim \d (\eta)$ due to
the terms containing second order derivatives w.r.t. $\eta$ (because of
$\pa_\eta^2 |\eta| = 2 \d (\eta)$), in other words, there
must be some {\em lightlike} ``thin shell'' matter present on the horizon.

\section{World-Volume Lagrangian Formulation of Lightlike Branes}

In a series of previous papers \ct{LL-brane-main,inflation-all,our-WH} we proposed
and studied manifestly reparametrization invariant world-volume actions describing
intrinsically lightlike $p$-branes for any world-volume dimension $(p+1)$:
\be
S = - \int d^{p+1}\s \,\P
\Bigl\lb \h \g^{ab} \pa_a X^{\m} \pa_b X^{\n} G_{\m\n}(X) - L\!\( F^2\)\Bigr\rb
\lab{LL-brane}
\ee
Here the following notions and notations are used:

\begin{itemize}
\item
Alternative non-Riemannian integration measure density $\P$ (volume form) on
the $p$-brane world-volume manifold:
\be
\P \equiv \frac{1}{(p+1)!}
\vareps^{a_1\ldots a_{p+1}} H_{a_1\ldots a_{p+1}}(B) \quad ,\quad
H_{a_1\ldots a_{p+1}}(B) = (p+1) \pa_{[a_1} B_{a_2\ldots a_{p+1}]}
\lab{mod-measure-p}
\ee
instead of the usual $\sqrt{-\g}$. Here $\vareps^{a_1\ldots a_{p+1}}$ is the
alternating symbol ($\vareps^{0 1\ldots p} = 1$), $\g_{ab}$ ($a,b=0,1,{\ldots},p$)
indicates the intrinsic Riemannian metric on the world-volume, and
$\g = \det\Vert\g_{ab}\Vert$.
$H_{a_1\ldots a_{p+1}}(B)$ denotes the field-strength of an auxiliary
world-volume antisymmetric tensor gauge field $B_{a_1\ldots a_{p}}$ of rank $p$.
Note that $\g_{ab}$ is {\em independent} of the auxiliary world-volume fields
$B_{a_1\ldots a_{p}}$. 
The alternative non-Riemannian volume form \rf{mod-measure-p} has been first
introduced in the context of modified standard (non-lightlike) string and
$p$-brane models in refs.\ct{mod-measure}.
\item
$X^\m (\s)$ are the $p$-brane embedding coordinates in the bulk
$D$-dimensional space-time with bulk Riemannian metric
$G_{\m\n}(X)$ with $\m,\n = 0,1,\ldots ,D-1$;
$(\s)\equiv \(\s^0 \equiv \t,\s^i\)$ with $i=1,\ldots ,p$;
$\pa_a \equiv \partder{}{\s^a}$.
\item
$g_{ab}$ is the induced metric:
\be
g_{ab} \equiv \pa_a X^{\m} \pa_b X^{\n} G_{\m\n}(X) \; ,
\lab{ind-metric}
\ee
which becomes {\em singular} on-shell (manifestation of the lightlike nature,
cf. Eq.\rf{gamma-eqs} below).
\item
$A_{a_1\ldots a_{p-1}}$ is an Auxiliary $(p-1)$-rank antisymmetric tensor gauge
field on the world-volume with $p$-rank field-strength and its dual:
\be
F_{a_1 \ldots a_{p}} = p \pa_{[a_1} A_{a_2\ldots a_{p}]} \quad ,\quad
F^{\ast a} = \frac{1}{p!} \frac{\vareps^{a a_1\ldots a_p}}{\sqrt{-\g}}
F_{a_1 \ldots a_{p}}  \; .
\lab{p-rank}
\ee
Its Lagrangian $L\!\( F^2\)$ is {\em arbitrary} function of $F^2$ with the
short-hand notation:
\be
F^2 \equiv F_{a_1 \ldots a_{p}} F_{b_1 \ldots b_{p}}
\g^{a_1 b_1} \ldots \g^{a_p b_p} \; .
\lab{F2-id}
\ee
\end{itemize}



Rewriting the action \rf{LL-brane} in the following equivalent form:
\be
S = - \int d^{p+1}\!\s \,\chi \sqrt{-\g}
\Bigl\lb \h \g^{ab} \pa_a X^{\m} \pa_b X^{\n} G_{\m\n}(X) - L\!\( F^2\)\Bigr\rb
\quad, \quad
\chi \equiv \frac{\P}{\sqrt{-\g}}
\lab{LL-brane-chi}
\ee
with $\P$ the same as in \rf{mod-measure-p},
we find that the composite field $\chi$ plays the role of a {\em dynamical
(variable) brane tension}.

Now let us consider the equations of motion corresponding to \rf{LL-brane}
w.r.t. $B_{a_1\ldots a_{p}}$:
\be
\pa_a \Bigl\lb \h \g^{cd} g_{cd} - L(F^2)\Bigr\rb = 0 \quad \longrightarrow \quad
\h \g^{cd} g_{cd} - L(F^2) = M  \; ,
\lab{phi-eqs}
\ee
where $M$ is an arbitrary integration constant. The equations of motion w.r.t.
$\g^{ab}$ read:
\be
\h g_{ab} - F^2 L^{\pr}(F^2) \llb\g_{ab}
- \frac{F^{*}_a F^{*}_b}{F^{*\, 2}}\rrb = 0  \; ,
\lab{gamma-eqs}
\ee
where $F^{*\, a}$ is the dual field strength \rf{p-rank}.

There are two important consequences of Eqs.\rf{phi-eqs}--\rf{gamma-eqs}.
Taking the trace in \rf{gamma-eqs} and comparing with \rf{phi-eqs}
implies the following crucial relation for the Lagrangian function $L\( F^2\)$:
\be
L\!\( F^2\) - p F^2 L^\pr\!\( F^2\) + M = 0 \; ,
\lab{L-eq}
\ee
which determines $F^2$ \rf{F2-id} on-shell as certain function of the integration
constant $M$ \rf{phi-eqs}, \textsl{i.e.}
\be
F^2 = F^2 (M) = \mathrm{const} \; .
\lab{F2-const}
\ee

The second and most profound consequence of Eqs.\rf{gamma-eqs} is that the induced
metric \rf{ind-metric} on the world-volume of the $p$-brane model \rf{LL-brane}
is {\em singular} on-shell (as opposed to the induced metric in the case of
ordinary Nambu-Goto branes):
\be
g_{ab}F^{*\, b}=0 \; ,
\lab{on-shell-singular}
\ee
\textsl{i.e.}, the tangent vector to the world-volume $F^{*\, a}\pa_a X^\m$
is {\em lightlike} w.r.t. metric of the embedding space-time.
Thus, we arrive at the following important conclusion: every point on the
surface of the $p$-brane \rf{LL-brane} moves with the speed of light
in a time-evolution along the vector-field $F^{\ast a}$ which justifies the
name {\em LL-brane} (lightlike brane) model for \rf{LL-brane}.

Before proceeding let us note that there exists a dynamically equivalent
dual Nambu-Goto-type world-volume action \ct{our-WH,inflation-all}
for the \textsl{LL-brane} producing the same equations of motion as the original
Polyakov-type \textsl{LL-brane} action \rf{LL-brane}:
\be
S_{\rm NG} = - \int d^{p+1}\s \, T
\sqrt{\bgv\, \det\Vert g_{ab} - \eps \frac{1}{T^2}\pa_a u \pa_b u\Vert\,\bgv}
\quad ,\quad \eps = \pm 1 \; ,
\lab{LL-action-NG-A}
\ee
where $g_{ab}$ indicates the induced metric on the world-volume \rf{ind-metric},
$u$ is the dual gauge potential w.r.t. $A_{a_1\ldots a_{p-1}}$
($F^{\ast}_{a} (A) = \mathrm{const}\,\chi^{-1}\pa_a u$),
and $T$ is {\em dynamical} tension simply proportional to the dynamical
tension in the Polyakov-type formulation \rf{LL-brane-chi}.

World-volume reparametrization invariance
allows to introduce the standard synchronous gauge-fixing conditions:
\be
\g^{0i} = 0 \;\; (i=1,\ldots,p) \; ,\; \g^{00} = -1
\lab{gauge-fix}
\ee
Also, we will use a natural ansatz for the ``electric'' part of the
auxiliary world-volume gauge field-strength:
\be
F^{\ast i}= 0 \;\; (i=1,{\ldots},p) \quad ,\quad \mathrm{i.e.} \;\;
F_{0 i_1 \ldots i_{p-1}} = 0 \; ,
\lab{F-ansatz}
\ee
meaning that we choose the lightlike direction in Eq.\rf{on-shell-singular}
to coincide with the brane
proper-time direction on the world-volume ($F^{*\, a}\pa_a \sim \pa_\t$).
The Bianchi identity ($\nabla_a F^{\ast\, a}=0$) together with
\rf{gauge-fix}--\rf{F-ansatz} and the definition for the dual field-strength
in \rf{p-rank} imply:
\be
\pa_0 \g^{(p)} = 0 \quad \mathrm{where}\;\; \g^{(p)} \equiv \det\Vert\g_{ij}\Vert \; .
\lab{gamma-p-0}
\ee
Then \textsl{LL-brane} equations \rf{gamma-eqs} acquire the form
(recall definition of $g_{ab}$ \rf{ind-metric}):
\be
g_{00}\equiv \Xdot^\m\!\! G_{\m\n}\!\! \Xdot^\n = 0 \quad ,\quad
g_{0i} = 0 \quad ,\quad g_{ij} - 2a_0\, \g_{ij} = 0
\lab{gamma-eqs-0}
\ee
(the latter are analogs of Virasoro constraints). Here $a_0$ is strictly positive
$M$-dependent constant:
\be
a_0 \equiv F^2 L^{\pr}\( F^2\)\bv_{F^2=F^2(M)} = \mathrm{const}
\lab{a-0}
\ee
($L^\pr(F^2)$ denotes derivative of $L(F^2)$ w.r.t. the argument $F^2$).
In particular, $a_0 = M$ for the ``wrong-sign'' Maxwell choice $L(F^2) = 1/4 F^2$.

Consider now codimension one \textsl{LL-brane} moving in a general
spherically symmetric background:
\be
ds^2 = - A(t,r)(dt)^2 + B (t,r) (dr)^2 + C(t,r) h_{ij}(\vec{\th}) d\th^i d\th^j \; ,
\lab{spherical-metric}
\ee
\textsl{i.e.}, $D=(p+1)+1$, with the simplest non-trivial ansatz for the
\textsl{LL-brane} embedding coordinates $X^\m (\s)$:
\be
t = \t \equiv \s^0 \;\; , \;\; r= r(\t) \;\; , \;\; \th^i = \s^i \;
(i=1,{\ldots},p) \; .
\lab{X-embed}
\ee
The \textsl{LL-brane} equations \rf{gamma-eqs-0},
taking into account \rf{gauge-fix}--\rf{F-ansatz}, acquire the form:
\br
-A + B \rdot^2 = 0 \;\; ,\; \mathrm{i.e.}\;\; \rdot = \pm \sqrt{\frac{A}{B}}
\quad ,\quad
\pa_t C + \rdot \pa_r C = 0
\lab{r-const}
\er
In particular, we are interested in static spherically symmetric metrics in
standard coordinates:
\be
ds^2 = - A(r)(dt)^2 + A^{-1}(r) (dr)^2 + r^2 h_{ij}(\vec{\th}) d\th^i d\th^j
\lab{standard-spherical}
\ee
for which Eqs.\rf{r-const} yield:
\be
\rdot = 0 \;\; ,\;\; \mathrm{i.e.}\;\; r(\t) = r_0 = \mathrm{const} \quad, \quad
A(r_0) = 0 \; .
\lab{horizon-standard}
\ee
Eq.\rf{horizon-standard} tells us that consistency of \textsl{LL-brane} dynamics in
a spherically symmetric gravitational background of codimension one requires the
latter to possess a horizon (at some $r = r_0$), which is automatically occupied
by the \textsl{LL-brane} (``horizon straddling'').

Similar feature (``horizon straddling'') occurs also for a codimension one
\textsl{LL-brane} moving in axially or cylindrically symmetric (rotating) backgrounds
\ct{our-WH,rot-WH}.
\section{Lightlike Brane as a Source of Einstein-Rosen ``Bridge'' Wormhole}

We will now show that the newly proposed metric
\rf{our-EF-metric}--\rf{our-EF-metric-coeff} is a
self-consistent solution of Einstein equations:
\be
R_{\m\n} - \h G_{\m\n} R = 8\pi T^{(brane)}_{\m\n}
\lab{Einstein-eqs}
\ee
derived from the action describing bulk ($D=4$) gravity coupled to an \textsl{LL-brane}:
\be
S = \int\!\! d^4 x\,\sqrt{-G}\,\frac{R(G)}{16\pi}\; +\; S_{\mathrm{LL}} \; ,
\lab{E-M-LL-4}
\ee
where $S_{\mathrm{LL}}$ is the \textsl{LL-brane} world-volume action \rf{LL-brane-chi}
with $p=2$.

Using the simplest non-trivial ansatz for the \textsl{LL-brane} embedding
coordinates $X^\m \equiv (v,\eta,\th,\vp)=X^\m (\s)$:
\be
v = \t \equiv \s^0 \;\; , \;\; \eta = \eta (\t) \;\; , \;\; \th^1 \equiv \th = \s^1
\;\;,\;\; \th^2 \equiv \vp = \s^2 \; ,
\lab{X-embed-0}
\ee
the pertinent \textsl{LL-brane} equations of motion yield:
\be
\eta (\t) = 0 \quad - \quad
\mathrm{horizon ~``straddling'' ~by ~the ~\textsl{LL-brane}} \; ,
\lab{eta-eq}
\ee
and the following expression for the \textsl{LL-brane} energy-momentum tensor:
\br
T_{(brane)}^{\m\n} = - \frac{2}{\sqrt{-G}}\, \funcder{S_{\mathrm{LL}}}{G_{\m\n}}
= S^{\m\n}\,\d (\eta)
\nonu \\
S^{\m\n} = \frac{\chi}{2a_0}\llb\pa_\t X^\m \pa_\t X^\n -
2a_0 G^{ij} \pa_i X^\m \pa_j X^\n \rrb_{v=\t,\,\eta=0,\,\th^i =\s^i} \; .
\lab{T-S-brane-0}
\er
Here $G^{ij}$ is the inverse metric in the $(\th^i) \equiv (\th,\vp)$ subspace
and $a_0$ indicates the integration constant parameter arising in the
\textsl{LL-brane} world-volume dynamics (Eq.\rf{a-0}).

Let us now turn to the Einstein equations \rf{Einstein-eqs} where we explicitly
separate the terms contributing to $\d$-function singularities on the l.h.s.,
\textsl{i.e.}, terms containing second-order derivatives w.r.t. $\eta$:
\be
R_{\m\n} \equiv \pa_\eta \G^{\eta}_{\m\n} - \pa_\m \pa_\n \ln \sqrt{-G} + \ldots
= 8\pi \Bigl( S_{\m\n} - \h G_{\m\n} S^{\l}_{\l}\Bigr) \d (\eta)
\lab{E-eqs-sing}
\ee
(the dots indicating non-singular terms).
Using the explicit expressions: 
\be
\G^{\eta}_{vv} = \h {\wti A}\pa_\eta {\wti A} \;\;,\;\;
\G^{\eta}_{v\,\eta} = -\h \pa_\eta {\wti A} \;\; ,\;\;
\G^{\eta}_{ij} = -\h {\wti A} G_{ij} \pa_\eta \ln {\wti r}^2
\;\; , \;\; \sqrt{-G} = {\wti r}^2
\lab{Christoffel-EF}
\ee
with ${\wti A} (\eta)$ and ${\wti r}(\eta)$ as in \rf{our-EF-metric-coeff},
taking into account ${\wti A}(0) = 0$ it is straightforward to check that non-zero
$\d$-function contributions in $R_{\m\n}$ appear for
$(\m\n)=(v\,\eta)$ and $(\m\n)=(\eta\eta)$ only. Substituting also the expressions
for the components of the \textsl{LL-brane} stress-energy tensor \rf{T-S-brane-0}
(with $G_{ij}$ indicating the metric in the $(\th^i) \equiv (\th,\vp)$ subspace) :
\be
S_{\eta\eta} = \frac{1}{2a_0}\chi \quad ,\quad S^\l_\l = - 2\chi \quad ,\quad
S_{ij} = - \chi\, G_{ij} \; ,
\lab{S-comp}
\ee
the Einstein equations \rf{E-eqs-sing} yield for
$(\m\n)=(v\,\eta)$ and $(\m\n)=(\eta\eta)$ the following matchings of the
coefficients in front of the $\d$-functions, respectively:
\be
m = \frac{1}{16\pi|\chi|} \quad ,\quad m = \frac{a_0}{2\pi|\chi|} \; .
\lab{M-Schw-EF}
\ee
where the \textsl{LL-brane} dynamical tension must be {\em negative}.
Consistency between the two relations \rf{M-Schw-EF} fixes the value
$a_0 = 1/8$ for the integration constant $a_0$ \rf{a-0}.
Most importantly, the first equation \rf{M-Schw-EF} shows that the mass parameter of
both Schwarzschild ``universes'' is determined uniquely by the dynamical
\textsl{LL-brane} tension.

From expressions \rf{S-comp} and the relation $S^\l_\l = 2\cP - \rho~$ the 
\textsl{LL-brane} pressure $\cP$ and energy density $\rho$ are identified to be:
\be
S_{ij} = \cP\, G_{ij} \;\; \to \;\; \cP = |\chi| \quad ,\quad \rho = 0 \; .
\lab{LL-pressure}
\ee
At this point let us note that violation of the null energy condition takes place 
(the \textsl{LL-brane} being an ``exotic matter'') as predicted by general wormhole 
arguments (cf. ref.\ct{visser-book}).

\section{Einstein-Rosen ``Bridge'' as a Limit of Spherically Symmetric
Wormhole with a Timelike Shell at the Throat}

Let us now consider a different modification of Eddington-Finkelstein form
of the Schwarzschild metric \rf{EF-metric} (cf.\rf{our-EF-metric}):
\be
ds^2 = - {\wti A}_1 (\eta) (dv)^2 + 2 dv\,d\eta +
{\wti r}_1^2(\eta) \llb (d\th)^2 + \sin^2\th (d\vp)^2\rrb \; ,
\lab{our-EF-metric-1}
\ee
where:
\be
{\wti A}_1 (\eta) = A (r_1 + |\eta|) \quad ,\quad
{\wti r}_1(\eta) = r_1 + |\eta| \quad ,\quad r_1 >2m \; ,
\lab{our-EF-metric-coeff-1}
\ee
\textsl{i.e.}, we now introduce a different change of coordinates from $r$ to
$\eta$ via $r=r_1 +|\eta|$.
The metric \rf{our-EF-metric-1}--\rf{our-EF-metric-coeff-1} describes two identical
copies of Schwarzschild {\em exterior}
space-time region ($r > r_1 (>2m)$), which correspond to
$\eta >0$ and $\eta <0$, respectively, and which are ``glued'' together at the
timelike hypersurface $\eta = 0$ (\textsl{i.e.}, $r=r_1 >2m$). The latter
hypersurface will serve as a ``throat''. As above
the metric \rf{our-EF-metric-1}  is smooth everywhere except at the
``throat'' $\eta = 0$ where it is only continuous but not differentiable.
Therefore, once again the Ricci tensor and the scalar curvature
will exhibit distributional contributions $\sim \d (\eta)$ due to the terms
containing second order derivatives w.r.t. $\eta$, in other words, they will
indicate the presence of a {\em timelike} ``thin shell'' matter on the throat.

Using Eqs.\rf{Christoffel-EF} for the current metric (\textsl{i.e.}, replacing
${\wti A}$ with ${\wti A}_1$ from \rf{our-EF-metric-coeff-1}) we find the
following results for the pertinent Ricci tensor components:
\br
R_{vv} = \frac{8m}{r_1^2} \Bigl( 1 - \frac{2m}{r_1}\Bigr)\,\d(\eta) \quad ,\quad
R_{v\eta} = - \frac{2m}{r_1^2}\,\d(\eta) \quad ,\quad
R_{\eta\eta} = - \frac{4}{r_1}\,\d(\eta)
\nonu \\
R_{ij} = - \frac{2G_{ij}}{r_1} \Bigl( 1 - \frac{2m}{r_1}\Bigr)\,\d(\eta)
\quad ,\quad R_{vi} = 0 \quad R_{\eta i} = 0  \; .
\lab{ricci-delta-1}
\er
Writing the Einstein equations for the metric
\rf{our-EF-metric-1}--\rf{our-EF-metric-coeff-1} in the form:
\be
R_{\m\n} = 8\pi \Bigl( S_{\m\n} - \h G_{\m\n} S^{\l}_{\l}\Bigr)\,\d (\eta)
\lab{Einstein-eqs-1}
\ee
and comparing with \rf{ricci-delta-1}, we identify the following timelike
``thin shell'' matter stress-energy tensor $S_{\m\n}$:
\br
S_{vv} = \frac{1}{2\pi r_1^2} \Bigl( 1 - \frac{2m}{r_1}\Bigr)\,
\Bigl(\frac{7m}{2}-r_1\Bigr) \quad ,\quad
S_{v\eta} = \frac{1}{2\pi r_1} \Bigl( 1 - \frac{2m}{r_1}\Bigr) \;
\nonu \\
S_{\eta\eta} = - \frac{1}{2\pi r_1} \quad ,\quad
S_{ij} = \frac{G_{ij}}{4\pi r_1} \Bigl( 1 - \frac{m}{r_1}\Bigr) \quad ,\quad
S_{vi} = 0 \quad, \quad S_{\eta i} = 0  \; .
\lab{S-comp-1}
\er
In the limit $r_1 \to 2m$  when the timelike ``thin shell'' (the throat) is moved
to the horizon, thus becoming a lightlike ``thin shell'', the only surviving
non-vanishing components of $S_{\m\n}$ read:
\be
S_{\eta\eta} = - \frac{1}{4\pi m} \quad ,\quad S_{ij} = \frac{G_{ij}}{16\pi m}
\quad ,\quad S^\l_\l = \frac{1}{8\pi m} \; .
\lab{S-comp-2}
\ee
Comparing Eqs.\rf{S-comp-2} with Eqs.\rf{S-comp} and accounting for
\rf{M-Schw-EF}, we see that the lightlike
limit of the ``thin shell'' matter at the throat coincides exactly with the
\textsl{LL-brane} with dynamical tension whose dynamics is consistently
described by the world-volume action $S_{\mathrm{LL}}$ \rf{LL-brane-chi}
appearing in \rf{E-M-LL-4}.
Yet, unlike the case with the \textsl{LL-brane}, the stress-energy tensor of
the timelike ``thin-shell'' (before the lightlike limit) {\em is not}
derived from any independent timelike world-volume brane Lagrangian.

Let us emphasize that we did not encounter any divergencies when taking the
lightlike limit $r_1 \to 2m$ above unlike the case with the usual procedure
for glueing together two outer Schwarzschild regions (with $r>r_1 (>2m)$)
along a timelike ``thin shell'' throat at $r=r_1$ (ref.\ct{visser-book},
Section 15.2.3).
The reason is that the Gaussian normal coordinate ${\bar \eta}$ used to describe
the normal direction w.r.t. hypersurface of the timelike ``thin shell'' turns
out to be {\em inappropriate} in the lightlike limit. Indeed, let us compare
the above construction of Einstein-Rosen-like wormhole with a {\em timelike} throat
given by the metric \rf{our-EF-metric-1}--\rf{our-EF-metric-coeff-1}
(Eqs.\rf{ricci-delta-1}--\rf{S-comp-2}) against
the standard construction using Gaussian normal coordinate ${\bar \eta}$
(cf. Eqs.(15.40)--(15.41) in ref.\ct{visser-book}):
\be
ds^2 = - {\bar A}({\bar \eta}) (dt)^2 + (d{\bar \eta})^2 +
{\bar r}^2 ({\bar \eta}) \llb (d\th)^2 + \sin^2\th (d\vp)^2\rrb \; ,
\lab{standard-gauss-normal}
\ee
where:
\be
{\bar A}({\bar \eta}) = 1 - \frac{2m}{{\bar r}({\bar \eta})} \quad ,\quad
\frac{d{\bar r}}{d{\bar \eta}} = \sqrt{{\bar A}({\bar \eta})} \; .
\lab{standard-gauss-normal-coeff}
\ee
The transformation between the coordinates $x\equiv (v,\eta,\th,\vp)$ and
${\bar x} \equiv (t,{\bar\eta},\th,\vp)$ relating the metrics \rf{our-EF-metric-1}
and \rf{standard-gauss-normal} is:
\br
t = v - r_1 - |\eta| -2m \ln \bv \frac{r_1 + |\eta|}{2m} -1\bv \;\; ,
\nonu \\
{\bar\eta} = {\bar\eta} (\eta) \quad \mathrm{where}\quad
\frac{d{\bar\eta}}{d\eta} = \frac{1}{\sqrt{{\wti A}(\eta)}}
\lab{coord-change}
\er
with ${\wti A}$ as in \rf{our-EF-metric-coeff-1}, and $\th,\,\vp$ -- unchanged.
Accordingly, the $D=4$ energy-momentum tensor of the ``thin shell'' 
transforms as $T_{\m\n}(x) =
{\bar T}_{\k\l}({\bar x})\,\partder{{\bar x}^\k}{x^\m}\,\partder{{\bar x}^\l}{x^\n}$
with $T_{\m\n}(x) = S_{\m\n}(v,\th,\vp)\,\d (\eta)$ and
\be
{\bar T}_{\m\n}({\bar x}) = {\bar S}_{\m\n}(t,\th,\vp)\,\d ({\bar\eta}) =
{\bar S}_{\m\n}(t,\th,\vp)\,\sqrt{1 - \frac{2m}{r_1}}\,\d (\eta) \; ,
\lab{T-S-transf}
\ee
where the last equation \rf{coord-change} has been used.
Therefore, we find for the pressure-defining parts of the ``thin shell''
stress-energy tensor (${\bar S}_{ij} = {\bar\cP}\, G_{ij}$ and
$S_{ij} = \cP\, G_{ij}$, respectively) following the coordinate transformation
relation:
\be
{\bar S}_{ij} = \Bigl( 1 - \frac{2m}{r_1}\Bigr)^{-\h}\, S_{ij}
\lab{Sbar-S}
\ee
with $S_{ij}$ -- the same as in \rf{S-comp-1}. Thus, for the
pressure ${\bar\cP}$ as defined within the standard approach using Gaussian
normal coordinate ${\bar\eta}$ we have:
\be
{\bar\cP} = \frac{1}{4\pi r_1} \frac{1-m/r_1}{\sqrt{1-2m/r_1}} \;\;
\longrightarrow \infty \;\;\; \mathrm{for}\;\; r_1 \to 2m ,
\lab{pressure-bar}
\ee
which exactly corresponds to Eq.(15.46) in ref.\ct{visser-book} and diverges
in the lightlike limit due to the prefactor on the r.h.s. of \rf{Sbar-S},
whereas for the pressure $\cP$ as defined within our choice of coordinates we get:
\be
\cP = {\bar\cP}\,\sqrt{1-2m/r_1} =
\frac{1}{4\pi r_1} \Bigl( 1 -\frac{m}{r_1}\Bigr) \;\; \longrightarrow
\frac{1}{16\pi\, m} = \mathrm{finite} \;\;\; \mathrm{for}\;\; r_1 \to 2m\; ,
\lab{pressure}
\ee
which precisely agrees with the second equation \rf{S-comp-2}
and with \rf{LL-pressure}--\rf{M-Schw-EF}. Accordingly, the ``thin-shell''
energy density: 
\be
\rho = {\bar\rho}\,\sqrt{1-2m/r_1} =
- \frac{1}{2\pi\, r_1}\bigl( 1 - \frac{2m}{r_1}\bigr)
\lab{e-density}
\ee
is negative and vanishes in the lightlike limit ($r_1 \to 2m$) in agreement 
with \rf{LL-pressure} (the expression for ${\bar\rho}$ in \rf{e-density} is the 
same as in the standard treatment using the metric \rf{standard-gauss-normal}, 
cf. Eq.(15.45) in ref.\ct{visser-book}).

The inappropriateness of the Gaussian normal coordinate ${\bar\eta}$ in the
lightlike limit can also be seen by observing that in the vicinity of the horizon
it coincides with the original Einstein-Rosen coordinate $u$ in \rf{E-R-metric}.

\section{Conclusions}

The original Einstein-Rosen ``bridge'' manifold \ct{einstein-rosen}, namely,
two identical copies of the outer Schwarzschild space-time region glued together
along the Schwarzschild horizon, appears as a particular case of the
general construction of spherically and axially symmetric traversable wormholes
produced by \textsl{LL-branes} as gravitational sources proposed in
refs.\ct{our-WH,rot-WH}. Here ``traversability'' means that a travelling observer
crosses the wormhole throat from the one ``universe'' to the other one within
a finite amount of his/her {\em proper} time. The same traversability property exists
also w.r.t. the Eddington-Finkelstein time $v$ (cf.\rf{our-EF-metric}).

The main lesson, as explained in some detail above, is that consistency of Einstein
equations of motion yielding the original Einstein-Rosen ``bridge'' as well-defined
solution necessarily requires the presence of \textsl{LL-brane}
energy-momentum tensor as a source on the right-hand side. Thus, the introduction
of \textsl{LL-brane} coupling to gravity brings the original Einstein-Rosen
construction in ref.\ct{einstein-rosen} to a consistent completion.

Codimension one \textsl{LL-branes} possess natural couplings to bulk Maxwell
$\cA_\m$ and Kalb-Ramond $\cA_{\m_1 \ldots \m_{p+1}}$ gauge fields ($D-1=p+1$,
see refs.\ct{LL-brane-main}):
\br
\wti{S}_{\mathrm{LL}} = \int d^{p+1}\s \,\P (\vp)
\llb - \h \g^{ab} \pa_a X^{\m} \pa_b X^{\n} G_{\m\n} (X) + L\!\( F^2\)\rrb
\nonu \\
-q \int d^{p+1}\s \vareps^{ab_1\ldots b_p} F_{b_1\ldots b_p} \pa_a X^\m \cA_\m (X)
\nonu \\
-\frac{\b}{(p+1)!} \int d^{p+1}\s \vareps^{a_1\ldots a_{p+1}}
\pa_{a_1} X^{\m_1} \ldots \pa_{a_{p+1}} X^{\m_{p+1}}\cA_{\m_1\ldots\m_{p+1}}(X)
\; ,
\lab{LL-brane-ext}
\er
where $q$ indicates the surface charge density of the \textsl{LL-brane}.
As shown in \ct{LL-brane-main}, the \textsl{LL-brane} can serve as a material
and charge source for gravity and electromagnetism by coupling it to bulk
Einstein-Maxwell+Kalb-Ramond-field system:
\be
S = \int\!\! d^D x\,\sqrt{-G}\,\llb \frac{R(G)}{16\pi}
- \frac{1}{4} \cF_{\m\n}\cF^{\m\n}
- \frac{1}{D! 2} \cF_{\m_1 \ldots\m_D}\cF^{\m_1 \ldots\m_D}\rrb
+ \wti{S}_{\mathrm{LL}} 
\lab{E-M-LL}
\ee
with $\wti{S}_{\mathrm{LL}}$ given by \rf{LL-brane-ext}.
Moreover, the \textsl{LL-brane} generates {\em dynamical} cosmological
constant through the coupling to the Kalb-Ramond bulk field: $\L = 4\pi \b^2$.


One can employ the above formalism to construct more general asymmetric
traversable wormholes, \textsl{e.g.}, glueing together an exterior
Schwarzschild region (``left'' universe) with an exterior Reissner-Nordstr{\"o}m
region  (``right'' universe) where the throat is the standard Schwarzschild horizon
w.r.t. the ``left'' universe and it is simultaneously the {\em outer}
Reissner-Nordstr{\"o}m horizon w.r.t. the ``right'' universe. Again as above
``traversability'' means traversability w.r.t. the proper time of travelling observers
crossing the throat.

\section*{Acknowledgments}
We are particularly indebted to Werner Israel and Eric Poisson for a series of
illuminating correspondence and incentive comments.
E.N. and S.P. are supported by Bulgarian NSF grant \textsl{DO 02-257}.
Also, all of us acknowledge support of our collaboration through the exchange
agreement between the Ben-Gurion University of the Negev (Beer-Sheva, Israel) and
the Bulgarian Academy of Sciences.


\end{document}